
\font\small=cmr9
\def\eqnum#1{\eqno (#1)}

\def\itm#1{\item{[#1]}}
\def\D#1{\displaystyle #1}
\def\PRL#1{{\sl Phys. Rev. Lett.} {\bf #1}}
\def\PRD#1{{\sl Phys. Rev.} {\bf D#1}}
\def\NPB#1{{\sl Nuc. Phys.} {\bf B#1}}
\def\PLB#1{{\sl Phys. Lett.} {\bf B#1}}
\def\vev#1{\langle #1 \rangle}
\def\SD{{\cal D}}
\def\nb{$n_{B}$}

\def\nbbar{$\bar{n}_{b}$}
\def\ng{$n_{\gamma }$}

\def\bin{$B_{in}$}
\def\mh{$m_H$}
\def\G{$\Gamma$}

\magnification=\magstep1
\hsize=6 true in
\vsize=8.5 true in
\baselineskip=19 true pt

\centerline{\bf Baryogenesis at the Electroweak
Scale\footnote{$^{\dag}$} {\small
Presented at Symposium on Early Universe, IIT, Madras, Dec.
1994. Work supported in part by the
Department of Science and Technology.}}
\bigskip
\centerline{U. A. Yajnik}
\centerline{\sl Physics Department, Indian Institute of Technology,}
\centerline{\sl Bombay 400\thinspace076}
\bigskip\bigskip
\centerline{\vbox{
\hsize=5 true in
\centerline{ABSTRACT}
\noindent
{\small The realisation that the electroweak anomaly can induce significant
baryon number violation at high temperature and that the standard
models of particle physics and cosmology contain all the ingredients
needed for baryogenesis has led to vigourous search for viable models.
The conclusions so far are that the Standard Model of particle physics
cannot produce baryon asymmetry of required magnitude. It has too
little $CP$ violation and sphaleronic transitions wipe out any
asymmetry produced if the Higgs is heavier than about 50 GeV, a range
already excluded by accelerator experiments. We review the sphaleron
solution, its connection to the high temperature anomalous rate and
then summarise possibilities where phenomenologically testable
extensions of the Standard Model may yet explain the baryon asymmetry
of the Universe.}
\vskip 1 true in}}

\noindent {\bf 1 Introduction}
\bigskip
An  observed  fact  of  nature  is  the  asymmetry  between   the
occurrence of matter and antimatter. This asymmetry is not of  a  local
nature as evidenced by an almost continuous distribution  of  luminous
bodies or hydrogen clouds in the Galaxy as also on  the  extragalactic
scales. Violent annihilation processes that may be  expected  at  the
boundary of regions containing matter and antimatter are also  not  to
be seen.[1] Since baryon number  is  a  good  symmetry  of  all  observed
processes, one has to assume the asymmetry to be of primordial  nature
and thus the problem passes into the domain of the early Universe.
\par
The net baryon number is not an important conserved number  from
the point of view of elementary particle physics. It is not  known  to
couple  to  any  gauge  bosons,  which  would   have   justified   its
conservation.
Thus  we  may  safely  assume  that  some  high  energy
processes yet undiscovered in fact violate baryon number[2]. Then in  the
fleeting moments of the early Universe, at  ultra  high  temperatures,
the number was not conserved and what we see is the residue left  over
after the annihilation of baryons and antibaryons.
The above line of thinking suffers from a further, deeper problem
when we compare the density of the  net  baryon  number $n_{B}$  with  the
entropy density $n_{\gamma }$ of photons. Standard Big Bang  cosmology
tells  us
that above  a  certain  temperature  in  the  early  Universe,  baryon
antibaryon pairs would be freely created by photons, and the approximate
thermodynamic equilibrium would make the  separate  densities  of  the
baryon number $n_{b}$ and the antibaryon number $\bar{n}_{b}$ to be of
the same  order
of magnitude as $n_{\gamma }$ aside from the asymmetry induced by processes
at an
even higher energy scale. If the high energy processes violated baryon
number freely, we would expect the asymmetry $n_{B}=n_{b}-\bar{n}_{b}$
to be of the same
order of magnitude as $n_{b}$ or $\bar{n}_{b}$. In that case, at
lower energies, after
the mutual annihilation, we expect $n_{b} = n_{B} \simeq  n_{\gamma }$.
Since both $n_{B}$ and $n_{\gamma }$
scale as $S^{-3}$, where $S$ is the Friedmann scale factor, the ratio of  the
two should  remain  constant  throughout  the  later  history  of  the
Universe.  This wishful thinking is contradicted by the observed
value   of  this  ratio[1],  which is  in   the   range
$10^{-10}-10^{-12}$. This is obtained by direct estimate  of  the  density  of
luminous matter and hydrogen clouds, compared to  entropy  density  of
the microwave background. There is a second method which confirms this
value, as well as  confirming  the  basic  premise  of  the  Big  Bang
cosmology. This is  the $n_{B}/n_{\gamma }$  ratio  needed
so  that  we  have  the
nucleosynthesis data about He to H by weight at its correct  value[1]  of
$25\%$. We are thus faced with  the  challenge  of  introducing  particle
physics interactions that violate baryon  number,  at  the  same  time
providing for the asymmetry, a fine tuned number in the range mentioned above.
\par
In the following,  in section 2,  we recapitulate  the  subtle
requirements for dynamical production of baryon number  in  the  early
Universe. In section 3, we introduce the sphaleron and the reasons for
looking for baryogenesis at the electroweak  scale.  We  also  discuss
here the bound placed on the Higgs boson mass by  the  requirement  of
electroweak baryogenesis.  In  section  4  we  introduce  a  class  of
mechanisms fulfilling the requirements of  electroweak  baryogenesis.
These will rely on the details of the electroweak phase transition, to
be  understood  in  terms  of  the  temperature  dependent   effective
potential. In section 5 we discuss the work done by our group  related
to previous  section,  viz.  baryogenesis in electroweak         phase
transition  induced  by  cosmic  strings.  Section  6   contains   the
conclusions. Due to limitations of space this review is  rather  brief
and selective. The references  cited  contain  more  details.  I  hope
however to convey the essentials and to share the excitement associated
with
making just enough baryons to ensure that we exist in the state we do.
\bigskip
\noindent {\bf 2 General Requirements}
\bigskip

The possibility of cosmological explanation of  baryon  asymmetry
relying on particle physics was proposed by Sakharov[3]
in the early days
of $CP$  violation  as  well  as  Microwave  Background.  Suppose  there
exist reactions in which baryon number (B)  is  violated.  However  if
there is charge conjugation symmetry (C), such reactions cannot  give  rise
to net baryon number, since both particles as  well  as  antiparticles
would be equally created or destroyed. Let us also suppose that $C$  is
violated in such reactions. However, as the reaction products begin to
build up , the reverse reactions would also become  viable,  returning
the products to the $B$ symmetric state. We must therefore also  suppose
existence  of  different  rates  for  the  forward  and  the   reverse
reactions, a possibility realisable in particle physics theories  with
violation of time reversal symmetry $T$, or  equivalently, $CP$ assuming
CPT invariance, $P$ being parity. However, in thermal
equilibrium, {\it CPT
invariance still implies equality of} $n_{b}${\it and } \nbbar{\it .}
Therefore  one  also
needs  out-of-equilibrium  conditions. In the   early
Universe, these could be provided by the decay or decoupling of
particles as certain temperature thresholds are crossed, or by
the occurance of a phase transition due to formation of condensates.
Then with the state of the Universe not time symmetric,  the
time irreversible processes could leave their distinct mark.
\par
It is clear that several factors have to conspire  rather
delicately to produce the required results. Our hope is that  we  make
hypotheses that are generic and yet lead to  this  rather  fine  tuned
number $n_{B}/n_{\gamma }$. Ideally we would like to allow for  maximal
possible $B$
violation as well as maximal possible $CP$ violation and have the  small
number come out compulsively as the result of a  distinct,  preferably
unique  mechanism.  Proposals  of  this  nature  were  made   in   the
context of grand unified theories, where out-of-equilibrium decays  of
superheavy bosons led naturally to the number needed.  At  present  we
have no compelling model of grand unified interactions,  but  that  is
not the reason why we shall turn to electroweak baryogenesis. It is in
fact deeper understanding of rather intricate facts  of  the  Standard
Model itself that lead us to look at  this  energy  scale  in  greater
detail.
\bigskip
\noindent {\bf 3 The anomaly and the sphaleron}
\bigskip

In Quantum Mechanics we usually expect the symmetries of the classical
system to be reflected as linear invariances of the Hibert space, and a
conserved quantity is expected to be represented by a hermitian
operator commuting with the Hamiltonian. This however is not always
true and a variety of other possibilities is now known to occur for the
case of Relativistic Field Theory. In the phenomenon known as
anomaly,[4] a classically conserved axial vector current associated
with fermions may turn out to be not conserved in the Quantum Theory.
Specifically, one finds that
\par
$$
\partial _{\mu }j^{\mu }_{A} = {g^{2}\over 32\pi ^{2}} \epsilon ^{\mu \nu \rho
\sigma }F_{\mu \nu }F_{\rho \sigma }
\eqnum{1}$$
\noindent where $g$ is the gauge coupling. This anomaly of the  fermionic
current
is associated with another interesting fact of gauge field theory.  It
was shown by Jackiw and Rebbi that the ground state of a    nonabelian
gauge thoery consists of many configurations  of  gauge  fields  which
although not permitting any nonzero physical field strengths, can  not
be continuously gauge transformed into each  other,  {\it i}.{\it e}.,  the
gauge
transformation connecting them cannot  be  deformed  to  the  identity
transformation.  Such  pure  gauge  vacuum   configurations   can   be
distiguished from each  other  by  a  topological  charge  called  the
Chern-Simons number
\par
$$
N_{C-S} = - {2\over 3}{g^3\over 32\pi^2} \int  d^{3}x \epsilon ^{\hbox{ijk}}
\epsilon ^{\hbox{abc}}A^{a}_{\nu }A^{b}_{\rho }A^{c}_{\sigma }
\eqnum{2}$$
The existence of gauge  equivalent  sectors  labelled  by  the
Chern-Simons number is related to the fermionic  anomaly  because  one
can show that the RHS of eqn.(1) is equal to a total divergence $\partial _{\mu
}K^{\mu }$
where
\par
$$
K^{\mu } = \epsilon ^{\mu \nu \rho \sigma } \pmatrix{F^{a}_{\nu
\rho } A^{a}_{\sigma }&-&{2\over 3}&g&\epsilon
^{\hbox{abc}}A^{a}_{\nu } A^{b}_{\rho }A^{c}_{\sigma }}
\eqnum{3}$$
\noindent so that with $F^{\mu \nu }=0,$
\par
$$
\Delta Q_{A} \equiv  \Delta  \int  j^{0}_{A} d^{3}x
= - \Delta \left( {g^{2}\over 32\pi
^{2}} \int K^{0} d^{3}x \right)
\equiv  - \Delta N_{C-S}
\eqnum{4}$$
\noindent
Thus the violation of the axial charge by unit occurs because of a
quantum transition from one pure gauge configuration to another.
The standard model  sphaleron[5] is  supposed  to   be   a   time
independent configuration of gauge  and  Higgs  fields  which  has
maximum energy along a minimal path joining sectors  differing  by
unit Chern-Simons  number. See figure 1.
\goodbreak
\midinsert
\vskip 2 true in
\centerline{{\bf fig. 1} Energy profile of gauge fields}
\endinsert

It  is  convenient  to  obtain
this  time independent solution in  the   approximation  that  the
Weinberg angle is zero.  Thus, the sphaleron of  an $SU(2)$  theory
spontaneously  broken  by  a  complex  isospinor  Higgs  is  given
by[5][6] (in the gauge $A^{a}_{0}= 0)$
\par
$$
\sigma ^{a}A^{a}_{i} = - {2i\over g} f(r){\partial
\over \partial x^{i}} U^{\infty }(\vec r)(U^{\infty }(\vec r))^{-1}
\eqnum{5}$$
\noindent and
\par
$$\phi (\vec r) = h(r)U^{\infty }(\vec r) \pmatrix{0\cr\mu }\eqnum{6}$$
\noindent with
\par
$$
U^{\infty }(\vec r) = {1\over r} \pmatrix{z&&x&+&iy\cr-x +&iy&&z&}
\eqnum{7}$$
\noindent The energy of such configurations can be estimated to
be $M_W/\alpha_{W} \simeq 10$ TeV. Detailed study[7][24] of the
sphaleron with  correct  value  of  the
Weinberg angle does not change the conclusions to be elaborated below.
The small mixing with the $U(1)_{Y}$ gives rise to a small magnetic  dipole
moment to the sphaleron, with accompanying modification in the energy.

In the  Standard  Model, $Q_{A}$
turns out to be the combination of the baryon and lepton numbers, $B+L$.
If we assume the physical vacuum to be  a  state  characterised  by  a
definite value of $N_{C-S}$, for instance the valley at
$N_{C-S}=0$ in fig. 1, a
spontaneous quantum transition  to another
state of $\Delta N_{C-S}=\pm 1$ can occur only by passing under
an energy barrier  of
height at least as much as set by the sphaleron energy
$E_{\hbox{sph}}\simeq 10$ TeV.
\par
Kuzmin, Rubakov and Shaposhnikov[8] conjectured
that  at  high  temperatures,  the  sphaleron  occurs   freely    as   a
fluctuation, and that the system  makes  transitions  to  neighbouring
valleys by going {\it over} the barrier. This would establish a
chemical equilibrium between  baryons
and antibaryons (as well as leptons and antileptons). Any  preexisting
asymmetry in the $B+L$ number  would  therefore  be  wiped  out  at  the
Weinberg-Salam phase transition scale.
\par
Subsequent analysis has substantiated this  conjecture
in two different temperature ranges: i$) 0 << T << E_{\hbox{sph}}$ and
ii$) T \ge  E_{\hbox{sph}}$ using  different  techniques.  Case  i)  is
amenable to reliable  approximation  techniques.[9] Accordingly,
the thermal rate for unit change in Chern-Simons number is
\par
$$
{\Gamma \over V} = {T^{4}\omega \over M_{W}(T)}
\pmatrix{{\alpha _{W}\over 4\pi }}^{4} N_{tr}N_{\hbox{rot}}
\pmatrix{{2M_{W}(T)\over \alpha _{W}T}}^{7}
\exp\{{-E_{\hbox{sph}}(T)\over kT}\} \kappa
\eqnum{8}$$
\noindent Here $\omega = \partial ^{2}V_{\hbox{eff}}/\partial
\phi ^{2}\mid _{\phi =0}; N_{tr}$ and $N_{\hbox{rot}}$ are
counts of sphaleron zero
modes  estimated  to  be $N_{tr}\times N_{\hbox{rot}}
\cong  1.3\times 10^{5}$  and $\kappa  {\char"7E} 1$ is  a
determinant. For case ii), no approximation techniques exist.
Sphaleron does not exist because $\langle \phi \rangle^T=0$ in
the high temperature regime.
But heuristic arguments suggest[10]
\par
$$
\Gamma  = A(\alpha _{W}T)^{4}
\eqnum{9}$$
\noindent where A is a dimensionless  constant  which  should  be  close  to
unity. We shall refer to this as the {\it high temperature} mechanism.
In order to establish this mechanism simulations have  been  carried
out on a lattice[11].  Gauge fields are set up on a lattice   in
contact with a heat bath and  allowed  to  evolve  in  fixed  time
steps, keeping track of the integral number $Q_{C-S}$  at every  stage.
The  one  such  calculation  carried   out[11] indeed   detects
occasional rapid jumps signalling $\Delta Q_{C-S}= \pm 1,$  and  an  empirical
value of A between 0.1 and 1.0. See fig. 2.
\goodbreak
\midinsert
\vskip 2 true in
\centerline{{\bf fig. 2} $N_{C-S}$ evolution in a thermal bath}
\endinsert

\par
$\Gamma $ is the anomalous transition rate ignoring  the  presence  of
fermions, and is equal for $\Delta Q_{C-S} = +1$ and
$\Delta Q_{C-S} = -1.$ The rate for
$B$-number violation is then obtained from  the  difference  between
the forward and reverse rates,  which  depend  upon  the  chemical
potentials of the baryons and the  antibaryons.   Then  the  final
result is
\par
$$
{\partial ( B + L)\over \partial t} = - {13\over 2} n_{f}{\Gamma \over T^{3}} (
B + L)
\eqnum{10}$$
\noindent where $n_{f}$  is  the  number  of  fermion  generations.  The
general
conclusion is therefore, that the anomaly is unsuppressed at  high
temperatures and no net $B + L$ can remain. This gives rise to
two broad options for explaining the baryon asymmetry: A) The
$B-L$ number of the Universe is non-zero for some reason, so
that with $B+L=0$, \nb$=n_L\neq0$ survives. For this to be a
satisfactory explanation, one needs a natural mechanism for
non-zero $n_{B-L}$. B) Net $B+L$ is generated at a scale not
much larger than the electroweak scale, and is neutralised
incompletely by high temperature electroweak processes. We shall
not take up a discussion of these possibilities but persue the
possibility of baryogenesis {\it at} the electroweak scale.
Hopefully, the latter approach has fewer free parameters and
will be easy to check against phenomenology.
\par
There is at least one important consequence of the above
analysis which we can derive within known phenomenology. Suppose
the net B-number just after the electroweak phase transition is
\bin. Since the rate in eq. (8) is known, we may integrate eq.
(10) to calculate how much B-number survives the menace of the sphaleron.
We note that the rate $\Gamma$ depends on the temperature
dependent expectation value $\langle\phi\rangle^T \equiv v$,
which in turn is determined by the parameters of the Higgs
potential, and hence in turn by the Higgs mass \mh. If the rate
\G\ is slow enogh to become comparable to the expansion rate of
the Universe, the sphalerons will not succeed in neutralising
the $B+L$ number. Following Shaposhnikov[12], we integrate eq.
(10) and obtain for suppression factor $S\equiv B_0/B_{in}$,
$$
S\ =\ \exp{(-\Gamma/H)}
\eqnum{11}$$
where $B_0$ is the net baryon number left over, and $H$ is the
Hubble parameter just after the electroweak scale. $S$ has an
implicit dependence on \mh, which is plotted in fig. 3.
\goodbreak
\midinsert
\vskip 2.5 true in
\centerline{{\bf fig. 3} Suppression factor vs. Higgs mass}
\endinsert
\noindent
We see that for large \mh\ such as \mh$>70$GeV, $B+L$ would
have to be zero {\it regardless of the physics beyond the
electroweak scale}. A light Higgs mass \mh$\sim 20$ to $35$ GeV
allows all the asymmetry before the electroweak phase transition
to survive. Assuming a modest value \bin$=10^{-5}$, we are led
to the conclusion \mh$<45$GeV since $B_0$ must be $10^{-10}$.
Sinc eaccelerator experiments have already ruled out \mh\
lighter than $57$GeV, this raises serious doubts about the
completeness of the Standard Model with one Higgs. This is
indeed the most significant result implied by the anomaly
structure of the Standard Model.
\bigskip
\noindent{\bf 4 Models of Electroweak Baryogenesis}
\bigskip

It is clear from the preceeding section that the Standard Model
needs to be modified, firstly to generate baryon asymmetry and
secondly, to prevent the wash out of the asymmetry. We shall
consider here some modifications to the Standard Model that are
minimal and satisfy the above requirements. In particular, we
dshall examine models in which the asymmetry is generated {\it
at} the electroweak phase transition.

As was explained in sec 2, time irreversible processes are an
essential ingredient of any recipe for baryogenesis. Kuzmin et
al[8] pointed out that this requirement could be met neturaly if
the electroweak phase transition was dirst order. Here by first
order we mean one in which the order parameter changes
discontinuously at the phase transition. The free energy of the
Higgs field is given in the field theoretic formalism by the
finite temperature effective potential. The high temperature
expansion for the same is given correct to $O(\hbar)$ by
$$
V^T_{eff}\ =\ -(2\lambda\sigma^2 + M_1^2 -
(M_2/\sigma)^2T^2)\phi^2 - {\D T\over4\pi}({\D M_3
\over\sigma})^3\phi^3
+ \lambda\phi^4 + ({\D M\over\sigma})^2\phi^4\ln ({\D
\phi\over\sigma})^2
\eqnum{12}$$
where $M_1$, $M_2$ and $M_3$ are mass dimension parameters
depending on physical masses $M_W$, $M_Z$, $M_t$; $\sigma$ is
the zero temperature expectation vallue of the Higgs field,
$\sigma=246$GeV; $\lambda$ determines the Higgs self-coupling.
The opposite signs between $T^2$ and the zero-temperature
coefficient in the $\phi^2$ term signals that for large enough
temperature, the effective mass-squared of the Higgs is positive
and the symmetry no longer appears broken[13]. The form of the
quartic potential leads to a variation in its shape with change
in the parameter T as shown in fig. 4.
\goodbreak
\midinsert
\vskip 3 true in
\centerline{{\bf fig. 4} Variation in $V^T_{eff}$ with $T$}
\endinsert
\noindent
There exists a temperature $T_1$ at which the system has two
equienergetic minima separated by a barrier. This barrier
persists till $T_c<T_1$, at which the second derivative of $V$
at $\phi=0$ changes sign from positive to negative so that
$\phi=0$ no longer remains a minimum. Between the temperatures
$T_1$ and $T_c$, the system is normally at $\phi=0$ since that
is the condition persisting from $T>T_1$. However since
$\phi_2\equiv\phi\neq0$ is favorable, thermal fluctuations and
quantum tunelling across the barrier is possible. Whenever
tunelling to the true vacuum occurs in any region of space, it
results in a ``bubble'', the inside of which consists of the
true vacuum $\phi_2$ and outside is still the unconverted false
vacuum. According to a well developed formalism[14][15], the
tunelling probability per unit volume per unit time is given by
$$
\gamma = CT^4e^{-S_{bubble}}
\eqnum{13}$$
where $S_{bubble}$ is value of
$$
S = 4\pi\int r^2dr\{{1\over2}\phi^{\prime 2} +
V^T_{eff}[\phi]\}
\eqnum{14}$$
extremised over $\phi$ configurations which satisfy the
``bubble'' boundary conditions $\phi(r=0)=\phi_2$, $\phi
\rightarrow 0$ as $r\rightarrow\infty$. Once a bubble forms,
energetics dictates that it keeps expanding, converting more of
the medium to the true vacuum. The expansion is irreversible and
provides one of the requisite conditions for producing
baryon asymmetry. The important question is whether the phase
transition in the electroweak theory is first order or second
order. Detailed calculations support the view that the form
of the potential is indeed as given above giving rise to a
temperature $T_1>T_c$, so that the phase transition is first
order. We thus have a B-number violating mechanism,
an irreversible process as well as the well known $CP$ violating
effects right within the Standard Model, thus giving rise to the
hopes of explaining the Baryon asymmetry at the electroweak scale.

Before proceeding with this discussion we should note that first
order phase transition with bubble formation is not the only way
time asymmetric conditions can arise. It has been pointed out by
Brandenberger and collaborators that even with a second order
phase transition, cosmic strings can play an important role in
catalising baryon asymmetry production if other favorable conditions
exist.[16] We can not include this interesting possibility for want
of space.

The hope expressed above of explaining B asymmetry within the
Standard Model is quickly belied by the fact that the extent of
known $CP$ violation is too small. A model independent
dimensionless parameter characterising the scale of this effect
has the value[17] $\delta_{CP}\sim 10^{-20}$. Since such a
factor is expected to appear multiplicatively in the final
answer, the resulting asymmetry would be too small.
Additianally, we saw that the Standard Model Higgs seems to
erase any B asymmetry generated prior to the electroweak scale.
This leads us to make the minimal extension to the Standard Model,
viz., to include one more complex Higgs doublet. The possibility
of such has been extensively considered in other contexts as
well[18]. For our purpose, this is a good extension to consider
for two reasons 1) a phase transition with two Higgs doublets
has the possibility of not wiping out the produced baryon
asymmetry and still allowing the lightest Higgs to be
heavier than 60 GeV[19]. 2) it is a source of additional $CP$ violation
which does not conflict with any known phenomenon.[18]

In the following we shall review one of the proposed scenarios
for electroweak baryogenesis in some detail, and refer to reader
to detailed reviews[20] for other possibilities. One class of
possibilities we are unable to take up is that due to Cohen
Kaplan and Nelson[21].

There are several proposals along these lines[20]. Unfortunately
we cannot include any details of most. One class of proposals by
Cohen Kaplan and Nelson involves scattering of neutrinos from
the walls of expanding bubbles. If the neutrino is massive and
has majorana mass, lepton number violation can occur in such a
scattering, biasing the $L$ number density in front of the wall,
after $CP$ violation has been taken into account. Outside the
wall, the high temperature anomalous process would be going full
swing, setting the $B+L$ number to zero, thereby creating
$B=-L$, i.e., negative of the $L$ generated by wall reflections.
On this basic scheme several phenomenologically viable models
have been proposed[21].

In the present review we shall treat in
some detail only one class of models proposed by McLerran,
Shaposhnikov, Turok and Voloshin[22]. Consider the model with
two Higgs doublets. In the course of the phase transition, both
of these acquire nonzero vacuum expectation value. Being
complex, their expectation values would generically differ in
their phase, thus allowing $CP$ violation in their nontrivial
ground state. The bias towards creation of baryons as against
antibaryons would be signalled by the presence of
terms in the effective action which are linear in the
Chern-Simmons number. Net baryon production can result only if
$CP$ violating effects are coupled to this biasing term.
A term with appropriate properties is contributed by the
triangle diagram shown in fig. 5.
\goodbreak
\midinsert
\vskip 2 true in
\centerline{{\bf fig. 5} A nontrivial contribution to the $S_{eff}$}
\endinsert
\noindent
The presence of two Higgs raises the danger of flavour changing
neutral currents, which is usually circumvented by coupling only
one of the Higgs to the fermions or coupling up type fermions to
one and down type to the other[23]. In either case, we get the
dominant contribution to above kind of diagram only from a top
quark loop with both scalar external legs coupled to the same
Higgs. The $T\neq0$ correction from this diagram can be
calculated to be
$$\eqalign{\Delta S={-7\over 4}\zeta(3)&\left({m_t\over \pi
T}\right)^2 {g\over16\pi^2}{1\over {v_1}^2}\cr
\times\int(\SD_i\phi_1^{\dag}\sigma^a
\SD_0 &\phi_1 - \SD_0\phi_1^{\dag}\sigma^a
\SD_i\phi_1)\epsilon^{ijk}F^a_{jk}d^4x\cr}\eqnum{15}$$
where $m_t$ is the top quark mass, $\zeta$ is the Riemann zeta
function, and the $\sigma^a$ are the Pauli matrices. For
homogeneous but time varying configurations of the Higgs fields,
in the gauge ${A_0}^a=0$, we can rewrite this in the
form
$$\eqalign{\Delta S\ =\ &{-i7\over4}\zeta(3)\left({m_t\over \pi
T}\right)^2 {2\over {v_1}^2}\cr
&\int dt [{\phi_1}^{\dag}\SD_0\phi_1 -
(\SD_0\phi_1)^{\dag}\phi_1]N_{CS}\cr
\equiv\ &{\cal O}N_{CS}\cr}
\eqnum{16}$$
which has the linear dependence on the $N_{CS}$ as required.
The expectation value of the operator $\cal O$, the imaginary part of
${\phi_1}^{\dag} \SD_0\phi_1$, acts as the
chemical potential for this number. $\langle \cal O \rangle$ is
nonzero only in the walls of bubbles, which is what we need.
However, to lowest adiabatic order Im$\phi_1$ can be made zero
by choice of gauge, and this persists when first derivatives are
taken. In the next adiabatic order, one finds
$$\eqalign{\SD_\mu \SD^\mu {\rm Im}\phi_1\ =\ {1\over2}R^3
\cos{\gamma} &(\lambda_5 \cos{\gamma} \sin{\gamma} \sin{\xi}\cr
&-\lambda_6 \sin^2\gamma \sin 2\xi)\cr}
\eqnum{17}$$
where $\vev{\phi_1}\sim\vev{\phi_2}\sim R$ in the translation
invariant ground state, $\gamma$ and $\xi$are angles specifying
relative phases of $\vev{\phi_1}$ and $\vev \phi_2$;
$\lambda_5$, $\lambda_6$ are dimensionless quartic couplings in
the two-Higgs doblet theory[18]. $\xi$ characterises the $CP$
violation which will show up only in the scalar sector. To
arrive at a numerical estimate, we take $\SD_\mu \SD^\mu \sim
M^2_H(T)$, the temperature dependent Higgs mass-squared, which
also sets the scale for the bubble wall thickness. This leads to[22]
\nb/\ng$\sim 10^{-3}\alpha_W^4 \sin 2\xi(T_c)$. If the quartic
couplings as well as $\sin 2\xi(T_c)$ are all $O(1)$, this leads
to an answer in the correct range of values.

The bubble profile can be computed by making reasonable ansatz  and the above
calculation can be done numerically. In a particular case of
bubble formation[24][28], one finds wall thickness
$\sim 40T^{-1}$ and  \nb/\ng\  indeed $\sim 10^{-9}$.
\bigskip
\noindent{\bf 5 String induced phase transition}
\bigskip

For all mechanisms relying on the first order nature of the
phase transition, the thickness and speed of the bubble walls
are crucial parameters. Some of the mechanisms would work only
in thin fast walls and others only for thick walls[20][25]. It
is possible that
the elecroweak phase transition was induced by cosmic strings
present from an earlier symmetry breaking transition. That this
is possible for a generic unified theory with several stages of
symetry breakdown was shown in ref. [26]. This was investigated
in detail for the electroweak effective potential in [27], where
it is shown that the thickness of bubble walls in this case is
$$\Delta r\ =\ s(m_H)T^{-1}
\eqnum{18}$$
where $s(m_h)$ is a scaling factor which varies in the range
$0.7-0.5\times(m_H/GeV)$ as $m_H$ varies from 60 to 120 GeV. For
the wall velocity we find $v\sim 0.5$ for the same range of
Higgs mass. A multiple time snapshot of the progressing bubble
wall is shown in fig. 6.
\goodbreak
\midinsert
\vskip 3.6 true in
\centerline{{\bf fig. 6} The induced bubble solution}
\endinsert

This mechanism invokes the existence of new gauge
forces at higher energies. However, the wall parameters givem
above are determined entirely by the standard model physics,
viz., $m_H$. These results show that the walls of string
induced bubbles provide adiabatic conditions for a B-asymmetry
generating mechanism, in particular conditions quite suited for
the operation of the McLerran-Shaposhnikov-Turok-Voloshin (MSTV)[22]
mechanism.

MSTV mechanism suffers from the drawback that to first adiabatic
order in the spatial variation of the Higgs fields, it does not
produce B-asymmetry. This happens because $CP$ violation comes
into play only if both the Higgs are involved whereas the FCNC
constraint forces coupling of the top quark on ly to one of the
two. Recently it has been shown[29] that the Glashow-Weinberg
criterion is sufficient but too strong, and that the possibility of
a fermion coupling to a small extent to another Higgs is open.
Accodingly we investigated[30] the MSTV mechanism with this
extension in the yukawa couplings. In this case two additonal
diagrams similar to the one in fig. 5 contribute to $S_{eff}$.
The value of the operator $\cal O$ was comuted in string induced
bubble walls. The results are numerically in the same
range; this is to be expected since the FCNC constraint still
keeps the contribution of additional diagrams small but the
effect is in the first adiabatic order, hence more robust.

\par
\noindent{\bf 6 Conclusion}
\par
For baryogenesis to occur in the early universe, three
conditions of Sakharov are necessary. In the Standard Model,
anomalous nature of the $B+L$ current allows for the violation
of this number. Further, the understanding of the sphaleron
solution permits the calculation of the rate of violation of
this number at high temperature, indicating that the rate of
violation becomes significant near the phase transition scale.
Numerical simulations also suggest that the vilation is
completely unsupressed above electroweak symmetry breaking
scale. Secondly, $CP$ violating interactions are possible in
simple extensions of the Standard Model, although the $CP$
violation is too small to produce the observed B-asymmetry. Finally,
upon investigating the
electroweak effective potential at high temperature, it is found
to suggest a first order phase transition, thus providing the
out-of-equilibrium conditions required by Sakharov's criteria.
This raises the possibility that all the observed baryon excess
in the Universe was manufactured at the electroweak scales and
mostly involving known physics. We have reviewed the MSTV[22]
mechanism involving two $SU(2)$ doublet scalars working as Higgs
particles. We find several variations of this mechanism that
would also be effective, in particular the one with less
stringent restriction on Yukawa couplings enhances this effect.

The highly effective baryon-number violation suggested by the
sphaleron and by numerical simulations above the electroweak
phase transition raise the spectre of a universe without baryons
if some mechanism for guarding them against the sphaleron menace
does not exist. Shaposnikov's work duely extended shows that
this implies that the Higgs mass in the Weinberg-Salam theory
must be less than about 50 GeV, a range already excluded by
accelerator experiments. This strongly suggests that the
Standard Model needs an extension. This is very valuable
information one derives about the fundamental forces at the
microscopic scale by studying the early Universe.

\bigskip
\noindent{\bf Acknowledgment}
\bigskip

I would like thank the organisers for the hospitality. The
research and travel were made possible by a DST sponsored project.

\vskip 1 true in\noindent
{\bf References}
\bigskip\noindent
\itm{1}
Kolb E. W. and Turner M. S., ``{\sl The Early Universe}'', Addison-Wesley
(1990)

\itm{2} Weinberg S., in ``{\sl Lectures on Particles and Fields}'', edited by
K.
Johnson at al, (Prentice-Hall, Englewood Cliffs, N.J., 1964),
pg. 482.

\itm{3} Sakharov A.D., {\sl JETP Lett.} {\bf 5}, 24 (1967)

\itm{4} See any Quantum Field Theory textbook, e.g., Huang K.,
``{\sl Quarks
Leptons and Gauge Fields}'', (World Scientific Pub. Co.,
Singapore).

\itm{5} Klinkhammer F. R. and Manton N. S., \PRD{30}, 2212 (1984)

\itm{6} See an early discussion in Polyakov A., {\sl Sov. Phys.-JETP},
{\bf 41}, 988 (1976)

\itm{7} Kleihaus B., Kunz J. and Brihaye Y., {\sl Phys. Lett}.
{\bf B}273, 100 (1992)

\itm{8} Kuzmin V. A., Rubakov V. A., and Shaposhnikov M. E.,
\PLB{155}, 36 (1985)

\itm{9} Arnold P. and McLerran L., \PRD{36}, 581 (1987); \PRD{37}, 1020 (1988)

\itm{10} Dine M., Lechtenfeld O., Sakita B., Fischler W., Polchinski J.,
\NPB{342}, 381,(1990)

\itm{11} Ambjorn J., Askgaard t., Porter H. and Shaposhnikov M.
E., \PLB{244},  479 (1990); \NPB{353}, 346 (1991)

\itm{12} Bochkarev A. I. and Shaposhnikov M. E., {\sl Mod. Phys.
Lett.} {\bf A2}, 417 (1987); Bochkarev A. I., Khlebnikov S. Yu.
and Shaposhnikov M.E., \NPB{329}, 490 (1990)

\itm{13} Kirzhnitz D. A. and Linde A. D., {\sl Sov. Phys.-JETP}
{\bf 40}, 628 (1974);\hfil\break
Dolan J. and Jackiw R., \PRD{9}, 3320 (1974);\hfil\break
Weinberg S., \PRD{9}, 3357 (1974)

\itm{14} Coleman S., \PRD{15}, 2929 (1977); Callan C. and
Coleman S., \PRD{16}, 1762 (1977).

\itm{15} Linde A. D., \PLB{70}, 306 (1977); \PLB{100}, 37 (1981);
\NPB{216}, 421 (1983)

\itm{16} Brandenberger R. and A-C. Davis, \PLB{308}, 79
(1993); Brandenberger R., Davis A-C. and Trodden M., \PLB{335},
123 (1994); Brandenberger  R. et al, preprint BROWN-HET-962.

\itm{17} Jarlskog C., \PRL{55}, 1039 (1985)

\itm{18} Gunion J. F., Haber H. E., Kane G. L. and Dawson S. ``{\sl The Higgs
Hunters Guide}'', (Addison-Wesley 1990)

\itm{19} Anderson G. E. and Hall L. J., \PRD{45}, 2685 (1992)

\itm{20} Cohen A., Kaplan D. and Nelson A., {\sl Ann. Rev. of
Nucl. and Particle Science} vol. {\bf 43}, 27 (1993)

\itm{21} Cohen A., Kaplan D. and Nelson A., {\sl Nuc. Phys.}\
{\bf B349}, 727 (1991); \PLB{263}, 86 (1991)

\itm{22} McLerran L., Shaposhnikov M.E., Turok N. and Voloshin
M. \PLB{256}, 351 (1991)

\itm{23} Glashow S. and Weinberg S., \PRD{15}, 1958  (1977)

\itm{24} Bhowmik Duari S., Ph.D. Thesis, IIT Bombay (1995), unpublished.

\itm{25} Dine M., Leigh R. L., Huet P., Linde A. and Linde D.,
\PLB{283}, 319 (1992); \PRD{46}, 550 (1992)

\itm{26} Yajnik U. A. \PRD{34}, 1237 (1986)

\itm{27} Bhowmik Duari S. and Yajnik U. A. \PLB{326}, 212 (1994)

\itm{28} Bhowmik Duari S. and Yajnik U. A., {\sl to appear in
Nucl. Phys. B, proceedings supplement, Workshop on Astroparticle
Physics, Stockholm University 1994}

\itm{29} Yu-Liang-Wu CMU-HEP93-19; DOE-ER/40682-44

\itm{30} Bhowmik Duari S. and Yajnik U. A., to be published.
See also ref. 24.
\bye